\documentclass[aps,onecolumn,prd,showpacs,showkeys,preprintnumbers,superscriptaddress,nobibnotes,notitlepage,floatfix,longbibliography,nofootinbib]{revtex4-2}

\usepackage{graphicx}
\usepackage{amsmath}
\usepackage{amssymb}
\usepackage{xspace}
\usepackage{xcolor}
\usepackage[normalem]{ulem} 

\usepackage[colorlinks=true,citecolor=blue,urlcolor=blue]{hyperref}

\def\be{\begin{equation}}
\def\ee{\end{equation}}
\def\bea{\begin{eqnarray}}
\def\eea{\end{eqnarray}}

\newcommand{\gsim}{\lower.7ex\hbox{$\;\stackrel{\textstyle>}{\sim}\;$}}
\newcommand{\lsim}{\lower.7ex\hbox{$\;\stackrel{\textstyle<}{\sim}\;$}}

\newcommand{\kpc}{\rm kpc}

\begin{document}

\title{How well do integral-of-motion distribution functions describe Milky Way-analogue halos?}

\author{Jason Kumar}
\affiliation{Department of Physics and Astronomy, University of Hawai'i, Honolulu, HI 96822, USA}

\author{Louis E.~Strigari}
\affiliation{
Mitchell Institute for Fundamental Physics and Astronomy,
Department of Physics and Astronomy, Texas A\&M University, College Station, TX 77843, USA
}

\begin{abstract}

We study the dark-matter phase space distribution of Milky Way-mass halos in the Auriga simulations, considering both hydrodynamic and dark-matter-only runs.  In a spherically averaged potential, we test whether the coarse-grained distribution function can be approximated as a function of the integrals of motion \(E\) and \(L\).  We find that \(f(E,L)\) provides a useful description of the simulated halos: although the deviations from this form are formally statistically significant, the median fractional residuals between radial shells are typically \(\lesssim 15\%\), with 90th-percentile residuals of order tens of percent.  An energy-only approximation, \(f(E)\), gives similar median residuals but is systematically disfavored by \(\chi^2\) tests, indicating that angular momentum dependence is statistically preferred.  Ultimately we find that neither \(f(E)\) nor \(f(E,L)\) fully describes the particle data at high precision.  The remaining residuals likely reflect departures from spherical equilibrium, including anisotropy, triaxiality, substructure, and baryonic effects.

\end{abstract}

\maketitle

\section{Introduction}

The dark matter (DM) phase space distribution is a key input to direct and indirect detection 
strategies.  Often, the velocity distribution is taken to obey a simple functional 
form, such as Maxwell-Boltzmann.  But if the phase space distribution is in equilibrium and 
obeys certain assumptions (such as spherical symmetry), then one can argue from principles 
of classical mechanics that the phase space distribution should be a function of only the 
energy ($E$) and angular momentum ($L$), implying a more complicated relationship between 
dark matter particle position and velocity.  Of course, these assumptions will never be 
exactly correct.  But if they are reasonably good approximations, then the phase space 
distribution can be well-approximated as a function of $E$ and $L$.  In this work, we will test 
how well this approximation is realized in publicly-available data from the Auriga~\cite{Grand:2024xnm} suite of 
dark matter halo numerical simulations.

We will consider Auriga runs containing only dark matter particles, and runs which include 
baryons (hydrodynamic).  For each run, we divide the halo into a large number of radial bins and determine 
the distributions of particles in $E$ and $L$ within each radial bin, to determine if they are 
consistent with a single function of $E$ and $L$, with no explicit dependence on $r$.  We will 
also be able to determine whether or not the phase space distribution is isotropic to a good approximation, 
with little dependence on $L$. 

In previous work~\cite{Christy:2023mgs}, a similar analysis was performed on 
publicly-available Via Lactea 2 data~\cite{Diemand:2008in}.  
In that case, it was found that the phase space distribution was statistically consistent with 
being a function of $E$ alone, well inside the scale radius of the halo.  
But in that study, statistical uncertainties were large  
because only $0.01\%$ of the Via Lactea 2 particle data was made publicly-available.  
In particular, there was not enough data to determine if the phase space distribution depended on 
$L$, let alone if there was a residual explicit dependence on $r$. 
However, all of the Auriga  data is publicly-available, allowing for a much more detailed 
study.  We find that the Auriga simulation phase space distributions are well approximated as 
being in equilibrium, and functions only of $E$ and $L$.  However, there are statistically 
significant deviations which indicate a residual explicit dependence on $r$.  But the 
typical spherically-averaged residuals are $\lesssim 15\%$.  
We will also find that, although the phase space distributions are typically anisotropic, the 
dependence of the phase space distribution  on $L$ is of limited importance.  Even if we approximate the 
phase space distribution as independent of $L$, the median spherically-averaged residuals are still 
 $\lesssim 15\%$.

If one approximates the phase space distribution as a function of $E$ alone, then it can be obtained 
from the density profile using the method of Eddington inversion~\cite{Eddington}.  Since the dark matter 
density profile can be estimated from stellar kinematics, one can then obtain the phase space 
distribution from data, allowing one 
to abstract 
these results beyond any particular numerical simulation run, and apply them to an arbitrary 
halo, including the Milky Way halo.  
This is especially important when considering indirect 
detection strategies when the dark matter annihilation cross section is velocity-dependent, 
in which case the annihilation rate depends on both the velocity and spatial 
distributions~\cite{Robertson:2009bh,Belotsky:2012zbe,Ferrer:2013cla,Boddy:2017vpe}.  
Knowing the velocity distribution at Earth is also important when considering the direct detection of dark matter~\cite{Vogelsberger:2008qb,Fox:2010bu}, particularly for models in which 
the energy deposited in a typical scatter is small, and one must take care in considering the 
velocity distribution in order to determine how many events deposit enough energy to exceed 
the experimental threshold.  
Indeed, it has been noted that the recent anomalous high energy event seen by the LZ experiment~\cite{LZ:2026axp}
may be the result of inelastic dark matter scattering, but this interpretation depends sensitively on 
the dark matter velocity distribution (see, for example, Ref.~\cite{Fan:2026kxx}).
Moreover, some strategies for the direct detection of dark matter 
rely on detecting an annual~\cite{Drukier:1986tm,Freese:2012xd} modulation in the event rate, 
due to the relative motion of the 
Sun with respect to the dark matter halo.  To properly model the modulation of the event rate, one must 
know the dark matter phase space distribution.  Our results will thus have impact on dark matter searches 
in a variety of contexts.

The plan of this paper is as follows.  In Section~\ref{sec:General_Formalism}, we review the 
general formalism for expressing the phase space distribution as a function of $E$ and $L$, and describe how we 
will determine if such a description is a good approximation to data from numerical simulations.  
In Section~\ref{sec:Results}, we apply 
that formalism to several Auriga halos, and present our results.  
We conclude in Section~\ref{sec:Conclusion}.

\section{General Formalism and Methods}
\label{sec:General_Formalism}

We consider the scenario in which the following assumptions hold:
\begin{itemize}
\item{The dark matter phase space distribution is static; that is, $f(\vec{x},\vec{p},t) = 
f(\vec{x},\vec{p})$.}
\item{The dark matter phase space distribution is spherically symmetric; that is, 
$f(\vec{x},\vec{p}) = f(r, p_r, p_\perp)$, where  $r=|\vec{x}|$ and where 
$p_r$ and $p_\perp$ are the radial and tangential components, respectively, of the 
momentum.}
\item{Each dark matter particle is subject only to central forces which depend on 
$r$.}
\end{itemize}
Then, since Liouville's Theorem\footnote{For a review, see Refs.~\cite{Goldstein,BinneyTremaine}.} states that $df/dt =0$ (and we assume $f$ has no explicit dependence 
on time), we must find that $f$ depends on $r$, $p_r$ and $p_\perp$ only through the six integrals of motion.
Since three of these integrals of motion determine the orientation of the particle's path in space (which is 
irrelevant if the phase space distribution is spherically-symmetric), and one more determines the particle's 
location along its path at any particular time (which is irrelevant if the phase space distribution is time-independent), 
the phase space distribution can only depend on the two remaining integrals of motion, namely $E$ and $L$.  This implies 
that $f$ depends on $r$, $p_r$ and $p_\perp$ only implicitly through the dependence of $E$ and $L$ on these variables.

If all of these assumptions were exactly true, then Liouville's Theorem would provide us with a great simplification, 
since $f$ would now depend on only two variables, instead of three.  In practice, one typically finds significant 
deviations from these assumptions in numerical simulations (see, for example, Ref.~\cite{Zemp:2009ff}), so one should expect 
deviations from the conclusion that the phase space distribution depends only on 
$E$ and $L$.  But if deviations from the assumptions are sufficiently small, then $f$ will be well-approximated 
as a function only of $E$ and $L$.  We will consider the extent to which this approximation holds in Auriga 
simulations.

Thus far, we have not made any assumptions regarding whether or not the DM phase space distribution is isotropic.  
If $f$ is isotropic, then its dependence on $p_r$ and $p_\perp$ can only be through the quantity $p = |\vec{p}| 
= \sqrt{p_r^2 + p_\perp^2}$.  We would then find that $f$ is independent of $L$, and depends only on $E$.  We will 
also consider the extent to which the approximation of an isotropic phase space distribution holds in 
Auriga simulations.

\subsection{Obtaining $f(E,L)$ and $f(E)$ from simulation data}

We will find it convenient to express the phase space distribution in terms of the momentum per unit mass (velocity) $v$.
If the phase space distribution is spherically symmetric, then the spatial density of particles per unit mass 
at any position $r$ can be expressed as 
\bea
\rho(r) &=& 2\pi \int_{-\infty}^\infty dv_r \int_0^\infty dv_\perp ~ v_\perp~f(r,v_r, v_\perp) ~,
\nonumber\\
&=& 2\sqrt{2} \pi \int_{\Phi(r)}^{\Phi(\infty)} dE \int_0^{\sqrt{2}r\sqrt{E-\Phi(r)}} dL ~ \frac{L}{r^2}
\frac{1}{\sqrt{E - \frac{L^2}{2r^2}-\Phi(r)}} f(E,L) ~,
\label{eq:rho}
\eea
where $L$ and $E$ are the angular momentum  and energy per mass,\footnote{Henceforth, we will frequently treat ``per mass" as implicit.} respectively, given by
\bea
L &=& v_\perp r ~,
\nonumber\\
E &=& \frac{1}{2} v_r^2 + \frac{L^2}{2r^2} + \Phi (r) ~.
\eea
$\Phi (r)$ is the gravitational potential, given by
\bea
\Phi (r) &=& \Phi (r_0) + G_N \int_{r_0}^r dx \frac{M(x)}{x^2} ~.
\label{eq:Phi}
\eea
Here, $M(r)$ is the mass enclosed within radius $r$, and $r_0$ is any convenient place to set the zero-point of the potential.  We will choose $r_0 =0$, $\Phi(0)=0$.  Note that the mass of the particles only enters in the definition 
of $\Phi (r)$.  If baryonic matter is included, then it also contributes to the enclosed mass.

Using eq.~\ref{eq:rho}, we can relate $f(E,L)$  
to the number of particles in a numerical simulation found within a 
radial shell, with energy and angular momentum found within particular bins.
Consider the particles lying within a radial bin (that is, a spherical shell) of size $\Delta V$ 
with mean radial coordinate $r_{avg}$.  Within this radial shell, we consider particles lying within 
an energy bin bounded by $E_{min}$ and $E_{max}$, and an angular momentum bin bounded by 
$L_{min}$ and $L_{max}$.
If we assume that the energy and angular momentum bins are small enough that $f(E,L)$ does not vary 
significantly over the size of the bin, then we can perform the energy, angular momentum and volume 
integrals in eq.~\ref{eq:rho} in order to relate $f(E,L)$ to $N(r,E,L)$, which is the number of particles in the 
simulation within a particular radial, energy and angular momentum bin.  We find
\bea
f(E,L) &=& \frac{N(r,E,L)}{\Delta V} \frac{3}{8\sqrt{2} \pi} \left[ 
\left(E_{max} - \frac{L_{min}^2}{2r_{avg}^2}-\Phi_{avg}(r) \right)^{3/2} 
- \left(E_{min} - \frac{L_{min}^2}{2r_{avg}^2}-\Phi_{avg}(r) \right)^{3/2}
\right.
\nonumber\\
&\,& \left.
-\left(E_{max} - \frac{L_{max}^2}{2r_{avg}^2}-\Phi_{avg}(r) \right)^{3/2}
+ \left(E_{min} - \frac{L_{max}^2}{2r_{avg}^2}-\Phi_{avg}(r) \right)^{3/2}
\right]^{-1} ~,
\label{eq:fEL_expression}
\eea
where $\Phi_{avg}(r)$ is the average value of $\Phi$ over particles within this radial bin.  
Note that, in expressing the integral in this form, we have assumed that the radial bins 
are small enough that we can treat $r$ as a constant over the radial bin with value $r_{avg}$, 
and can treat $\Phi(r)$ as a constant over the radial bin with value $\Phi_{avg}(r)$.  Moreover, 
we have assumed that $f(E,L)$ can be treated as a constant over the energy and angular momentum 
bins.  However, this is only an approximation, and introduces an error which increases with the 
size of the bins.  In particular, for sufficiently large bins and for values of $E_{min,max}$ and 
$L_{min,max}$ near the boundary of the kinematically allowed range, the right-hand-side of 
eq.~\ref{eq:fEL_expression}  may not be real.

Eq.~\ref{eq:fEL_expression} provides an expression for $f(E,L)$ derived from every 
radial bin for which those values of $E$ and $L$ are kinematically accessible (up to the binning error 
mentioned above).  
If the phase space 
distribution is, in fact, only a function of $E$ and $L$, then the values of $f(E,L)$ derived from every 
such radial bin should be the same, up to the 
statistical fluctuations inherent in a numerical simulation (and up to the inherent binning 
error).  On the other hand, if the values of 
$f(E,L)$ derived from different radial bins are not consistent with each other, this would indicate 
a residual explicit dependence on $r$ (that is, $f=f(E,L,r)$), providing a measure of the deviation of 
the numerical simulation from our assumptions of a static, spherically-symmetric phase space distribution.

If the phase space distribution is also isotropic (that is, if $f(E,L) = f(E)$), then the integral over 
$L$ in eq.~\ref{eq:rho} can be performed analytically, yielding 
\bea
\rho(r) &=&  4\sqrt{2} \pi \int_{\Phi(r)}^{\Phi(\infty)} dE~ f(E) 
\sqrt{E -\Phi(r)} ~.
\label{eq:fE}
\eea
Proceeding as above, we can relate $f(E)$ to the number of particles found within a particular 
radial shell and energy bin ($N(r,E)$), yielding 
\bea
f(E) &=& \frac{N(r,E)}{ \Delta V} \left[ \frac{8\sqrt{2} \pi}{3}  \left([E_{max} -\Phi_{avg}(r)]^{3/2} - [E_{min} -\Phi_{avg}(r)]^{3/2}\right) \right]^{-1} ~.
\label{eq:fE_expression}
\eea

\subsection{Statistical Analysis}

To determine if the phase space distribution is well-described by a function of $E$ and $L$ alone, 
we define $f_i (E_j,L_k)$ as the value of 
$f(E_j,L_k)$ obtained from the $i$th radial bin (with radius $r_i$), as given in eq.~\ref{eq:fEL_expression}.  
Here, $E_j$ represents the energy 
of the $j$th energy bin, and $L_k$ represents the angular momentum of the $k$th angular momentum bin.

We only consider radial, energy and angular momentum bin triplets $(i,j,k)$ such that  the terms in eq.~\ref{eq:fEL_expression} are well defined and such that $N(r,E,L) > 20 $.  The latter requirement 
ensures that the bin has adequate statistical power.
We will denote by ${\cal B}$ the set of such bin triplets.  The number of such triplets is 
$N_{bins}$.

The fractional statistical uncertainty 
in $f_i (E_j, L_k)$ is then 
\bea
\frac{\delta f_i (E_j, L_k)}{f_i (E_j, L_k)} &=& \frac{1}{\sqrt{N(r_i,E_j,L_k)}} ,
\label{eq:fE_uncertainty}
\eea
where we have assumed Gaussian statistics.  Since we only consider bin triplets for which 
$N(r,E,L) > 20 $, this will be a good approximation.

We can then obtain a weighted average of $f(E_j,L_k)$ over all relevant radial bins.  The  
weighted average $\bar{f}$ (along with its uncertainty, $\delta \bar{f}$), are defined as
\bea
\bar{f} (E_j, L_k) &=& \frac{\sum_i f_i (E_j, L_k) ~\left(\delta f_i (E_j, L_k) \right)^{-2} }
{\sum_i \left(\delta f_i (E_j, L_k) \right)^{-2}} ,
\nonumber\\
\delta \bar{f} (E_j, L_k) &=& \left[ \sum_i \left(\delta f_i (E_j, L_k) \right)^{-2}  \right]^{-1/2} ,
\label{eq:f_avg}
\eea
where the sum is over radial bins $i$ such $(i,j,k) \in {\cal B}$.  
We will let $N_f$ denote the number of energy and angular momentum bin doublets $(j,k)$ such that 
$\bar{f}(E_j, E_k)$ is well-defined (that is, there is at least one term in the 
sum in eq.~\ref{eq:f_avg}).

To measure the consistency of the simulation data in all radial bins with the phase space distribution given by 
$\bar{f} (E,L)$, we define 
\bea
\chi^2 &=& \sum_{(i,j,k) \in {\cal B}} 
\frac{\left(f_i (E_j, L_k) - \bar{f}(E_j, L_k) \right)^2}{\delta f_i^2 (E_j,L_k)   } .
\eea
The  number of degrees of freedom of the $\chi^2$-distribution is $N_{dof} = N_{bins} - N_{f}$.

If we approximate the phase space distribution as isotropic, then we can obtain 
$f_i(E_j)$ from simulation data using eq.~\ref{eq:fE_expression}.  We can then 
perform a similar statistical analysis to find $\bar{f} (E_j)$, $\delta \bar{f} (E_j)$ 
and $\chi^2$ by assuming a single angular momentum bin.  

\section{Results}
\label{sec:Results}

We will apply this analysis to several members of the Auriga suite of simulations~\cite{Grand:2016vxo,Grand:2024vuj}. The Auriga simulations are a suite of cosmological magnetohydrodynamic zoom-in simulations of Milky Way-mass halos, run with the moving-mesh code \textsc{Arepo} and selected to have relatively isolated assembly histories. We use the publicly released level-4 ``Original'' simulations at redshift \(z=0\), considering halos 6, 9, and 16. For each object we analyze both the hydrodynamic run, which includes gas dynamics, star formation, stellar feedback, black-hole physics, and the associated baryonic contribution to the gravitational potential, and the matched dark-matter-only run. The virial radii of the halos considered here are \(R_{200}\simeq 214\)--\(241\,{\rm kpc}\), and the dark-matter particle masses are of order \(2\times 10^5\,M_\odot\). 
The radius of convergence is taken to be $1.3~\kpc$~\cite{Board:2021bwj}.
Throughout this work the phase space distribution is estimated using only dark-matter particles; in the hydrodynamic runs, however, the potential used to define particle energies is computed from the total enclosed mass, including baryons. In each case, 
we apply our analysis to the last timestamp.
As an initial step, we will test the extent to which the simulation data satisfies the assumptions of 
spherical-symmetry, equilibrium, and isotropy.

\subsection{Spherical Symmetry}

We first locate the center of the halo using the shrinking sphere method of Ref.~\cite{Board:2021bwj}.
Having shifted the center of mass to the origin, spherical symmetry is a good approximation~\cite{Board:2021bwj}.
We can quantify deviations from spherical symmetry by expanding the spatial density  in spherical harmonics.
The spherical harmonic coefficients are defined as 
\bea
a_{\ell m} &=& \frac{\int dV \rho(r, \theta, \phi) ~ Y_{\ell m} (\theta, \phi)} {\int dV \rho(r, \theta, \phi) } 
= \frac{1}{N} \sum_{i=1}^N Y_{\ell m} (\theta_i, \phi_i) ,
\eea
where we have taken  $\rho (r, \theta, \phi) = \sum_{i=1}^N (1 / r_i^2)\delta(r-r_i) \delta (\cos \theta - \cos \theta_i) \delta (\phi - \phi_i)$.
The sum is over only the $N$ dark matter particles outside the radius of convergence and  inside $R_{200}$. 
For the $i$th such particle, the position of the particle may be expressed as 
$x_i = r_i \sin \theta_i \cos \phi_i$, $y_i = r_i \sin \theta_i \sin \phi_i$, $z_i = r_i \cos \theta_i$.

Computing the $a_{\ell m}$ for $\ell \leq 2$, we  present the maximum value of 
$|a_{\ell m}|/a_{00}$ for each halo in Table~\ref{tab:halo_parameters}.  From these 
results, we see that spherical symmetry is a good approximation, but by no means 
exact.

\subsection{Equilibration}

To quantify the extent to which the halo is in equilibrium, we can use the virial theorem (after subtracting 
off the bulk velocity of the halo).   
We should have 2$ \sum_i \overline{T_i}  = -  \sum_i \overline{\vec{F}_i \cdot \vec{r}_i} $, 
where $\overline{X }$ means the time-average of $X$ over a long time and the sum is over all DM particles inside 
$R_{200}$.  
If the DM distribution is in equilibrium, then the time 
average of the sum will be well approximated by evaluating the sum using our data from the last time step.  We define
\bea
q &\equiv& \frac{\sum_j v_j^2}{\sum_i G_N M(r_i) / r_i} = \frac{\langle v^2 \rangle}{\langle G_N M(r) / r \rangle} ,
\nonumber\\
\delta q &=& \frac{\langle v^2 \rangle}{\langle G_N M(r) / r \rangle} 
\left[\frac{\langle v^4 \rangle - \langle v^2 \rangle^2 }{N\langle v^2 \rangle^2} 
+ \frac{\langle (M(r) / r)^2 \rangle - \langle M(r) / r \rangle^2 }{N\langle M(r) / r \rangle^2}\right]^{1/2} ,
\eea
where $\langle ... \rangle$ is the average over the $N$ particles, and $M(r)$ is the mass enclosed within 
a radius $r$.
$\delta q$ is obtained by propagation of errors.
We can then measure how well the approximation of a time-invariant distribution holds by seeing how close $q$ is to 1.

We present the values of $q$ obtained for each halo in Table~\ref{tab:halo_parameters} (the 
statistical uncertainties are negligible).  We thus see that there are statistically significant 
deviations from equilibrium, but only at the $\lesssim 10\%$ level, indicating 
that a static distribution is a very good approximation, though not exact. 

\begin{table}
    \centering
    \begin{tabular}{|c|c|c|c|c|c|}
    \hline
      Halo & baryons? & $\gamma$ & $|a_{1m}^{max}|/a_{00}$ & $|a_{2m}^{max}|/a_{00}$  & $q$  \\
    \hline
      6 & Yes & $1.495$ & $0.09$ & $0.19$ & $1.08 $  \\
      6 & No & $1.198$ & $0.10$ & $0.22$ & $1.11$ \\
      9 & Yes & $1.793$ & $0.03$ & $0.071$ & $1.03$ \\
      9 & No & $1.226$ & $0.02$ & $0.22$ & $1.07$ \\
      16 & Yes & $1.250$ & $0.12$ & $0.15$ &   $1.12 $ \\
      16 & No & $1.134$ & $0.14$ & $0.19$ & $1.13$ \\
    \hline
    \end{tabular}
    \caption{Table of halos considered, $\gamma$, $|a_{1m}^{max}|/a_{00}$, $|a_{2m}^{max}|/a_{00}$ and 
    $q$, as described in the text.}
    \label{tab:halo_parameters}
\end{table}

\subsection{Anisotropy}

We investigate how close phase space distributions are to isotropic in any radial shell, labeled by $i$.  
To do this, we define a spherical aniostropy parameter $\beta_i$ by
\bea
\beta_i &\equiv& 
1 - \frac{\langle v_\perp^2 \rangle}{2\langle v_r^2 \rangle},
\nonumber\\
\delta \beta_i &=&  \frac{\langle v_\perp^2 \rangle}{2\langle v_r^2 \rangle} 
\left[\frac{\langle v_r^4 \rangle - \langle v_r^2 \rangle^2}{N_j \langle v_r^2 \rangle^2} 
+ \frac{\langle v_\perp^4 \rangle - \langle v_\perp^2 \rangle^2}{N_j \langle v_\perp^2 \rangle^2} \right]^{1/2} ,
\eea
where the average is over all $N_i$ DM particles in the radial shell.  
$\delta \beta_i$ is obtained by propagation of errors.
If the velocity distribution in this shell is 
close to isotropic, then $\beta_i$ will be close to 0.  We can thus use $\beta_i$ to 
investigate deviations from isotropy as a 
function of $r$.  We can restrict ourselves to $r < R_{200}$.

\begin{figure}
    \centering
    \includegraphics[width=1.0\linewidth]{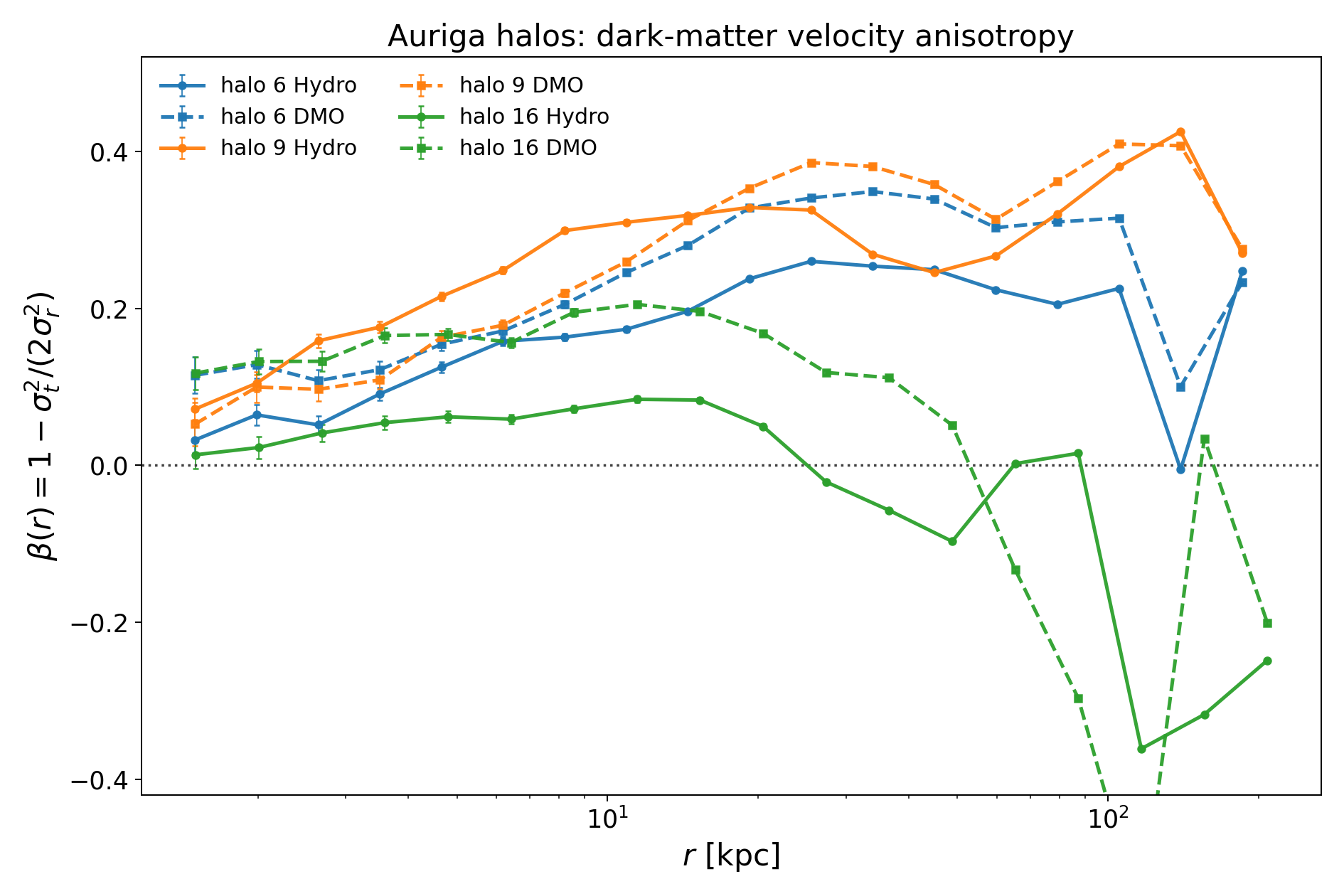}
    \caption{Spherical velocity anisotropy profiles for the Auriga halos analyzed in this work. The quantity \(\beta(r)=1-\sigma_t^2/(2\sigma_r^2)\) is computed in radial shells using bound dark-matter particles between the radius of convergence and \(R_{200}\). Hydrodynamic and dark-matter-only runs are shown for halos 6, 9, and 16. The profiles show significant departures from isotropy, particularly in the outer halo, motivating the comparison between isotropic \(f(E)\) and anisotropic \(f(E,L)\) descriptions of the phase-space distribution.}
    \label{fig:Figure1}
\end{figure}

In Figure~\ref{fig:Figure1}, we plot $\beta_i$ for each radial shell, for each halo which we consider.  
The uncertainties are shown with error bars; 
they are negligible at large $r$.
We find that there are significant deviations from isotropy, which tend to be larger far from the center 
of the halo.  As expected~\cite{Zemp:2009ff}, we find that isotropy is a better approximation 
near the center of the halo.

\subsection{Phase space distribution}

For each halo, we fit the density profile for all dark matter particles between the radius of 
convergence and $R_{200}$ to a generalized NFW (gNFW) profile of the form 
$\rho(r) = \rho_s (r/r_s)^{-\gamma} [1 + (r/r_s)]^{\gamma-3}$, where $\rho_s$ and $r_s$ are 
the scale density and scale radius, respectively.  The inner slope parameter $\gamma$ for each 
halo is presented in Table~\ref{tab:halo_parameters}.

We have seen that the Auriga simulation halos seem to be well approximated as spherically symmetric and 
equilibrated.  However, significant anisotropy is found in the velocity distribution.  This implies that 
$f$ may be well-approximated as a function of both $E$ and $L$, and the dependence on $L$ may be non-trivial.

To test these conclusions, we divide each halo into $1000$ radial bins, $64$ energy bins, and 
$64$ angular momentum bins. 
The radial bins are evenly spaced logarithmically, from the radius of convergence to $R_{200}$.\footnote{Note that, in computing the gravitational potential, all 
bound particles are used, including those within the radius of convergence.}  The energy 
bins are linearly spaced between the 2nd and 98th percentile of the energies of bound particles, and 
the angular momentum bins are linearly spaced between 0 and the 98th percentile of particles bound to the halo.

\begin{figure}
    \centering
    \includegraphics[width=1.0\linewidth]{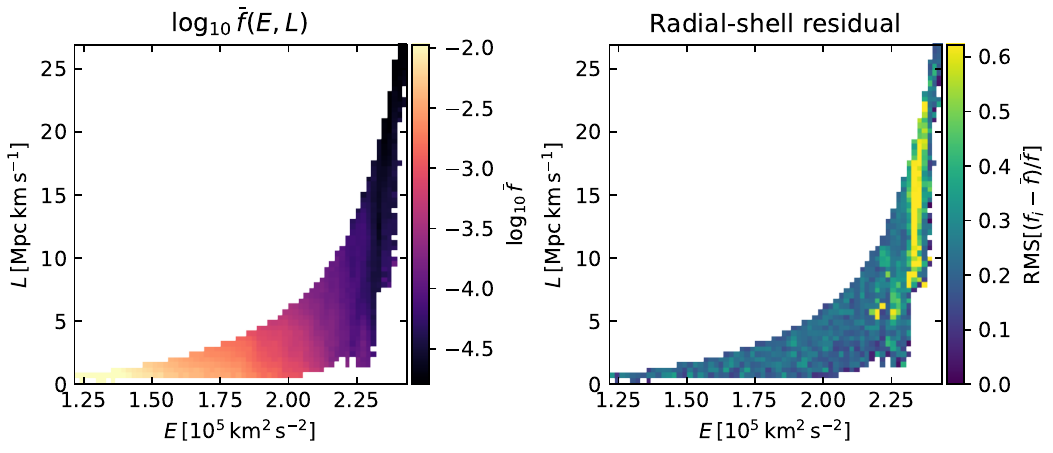}
    \caption{Inferred distribution function and residuals for the Auriga halo 6 hydrodynamic run. Left: weighted shell-averaged distribution function \(\log_{10}\bar f(E,L)\), obtained by combining estimates from radial shells. Right: RMS fractional residual between the individual radial-shell estimates \(f_i(E,L)\) and the shell-averaged value \(\bar f(E,L)\). Empty regions are kinematically inaccessible or insufficiently populated. 
       }
    \label{fig:fEL}
\end{figure}

We use the simulated data to determine $f(E,L)$ in each radial bin and determine $\bar{f} (E,L)$ 
using the methods described in Section~\ref{sec:General_Formalism}.  
Figure~\ref{fig:fEL} shows the resulting shell-averaged distribution function for one representative halo, together with the radial-shell residuals about this average. The figure illustrates both the smooth leading dependence of the coarse-grained phase-space distribution on \(E\) and \(L\), and the residual shell-to-shell structure that remains after averaging over radius.  Note that the residuals tend to be largest at large energy, where the phase-space distribution describes particles which can access a large range of radii.

In Table~\ref{tab:FEL_results}, we 
present $N^{E,L}_{dof}$ and $\chi_{E,L}^2/dof$ for each halo.  
This $\chi^2$-distribution characterizes the statistical significance with which the actual phase space distribution
deviates from a function of $E$ and $L$ alone.
We also present the median fractional residual, which 
is defined as 
\bea
(|\Delta f|/f)_{med.}^{E,L} &=& \mathrm{median} \left(\frac{|f_i(E_j,L_k) - \bar{f}(E_j,L_k)|}{\bar{f}(E_j,L_k)} \right),
\eea
where the median is taken over all $(i,j,k) \in {\cal B}$.  $(|\Delta f|/f)_{med.}$ describes the 
extent to which $f_i (E_j, L_k)$ obtained in any radial bin differs from $r$-independent phase space 
distribution $\bar{f}(E_j, L_k)$, and thus characterizes 
the fractional error made in approximating the phase space distribution as a function of 
$E$ and $L$ alone.

For typical halos, we find  $\chi_{E,L}^2/dof \sim 1-2$ and  $N^{E,L}_{dof }\sim 30000-40000$.    Given the large number of degrees of freedom, the resulting values of $\chi_{E,L}^2/dof \sim 1-2$ 
indicate that there are strong statistical deviations 
from the assumption that the phase space distribution is a function of $E$ and $L$ alone.  However, we also 
see that the median fractional residuals are $\sim 15\%$, characterizing the magnitude of these deviations.

\begin{figure}
    \centering
    \includegraphics[width=1.0\linewidth]{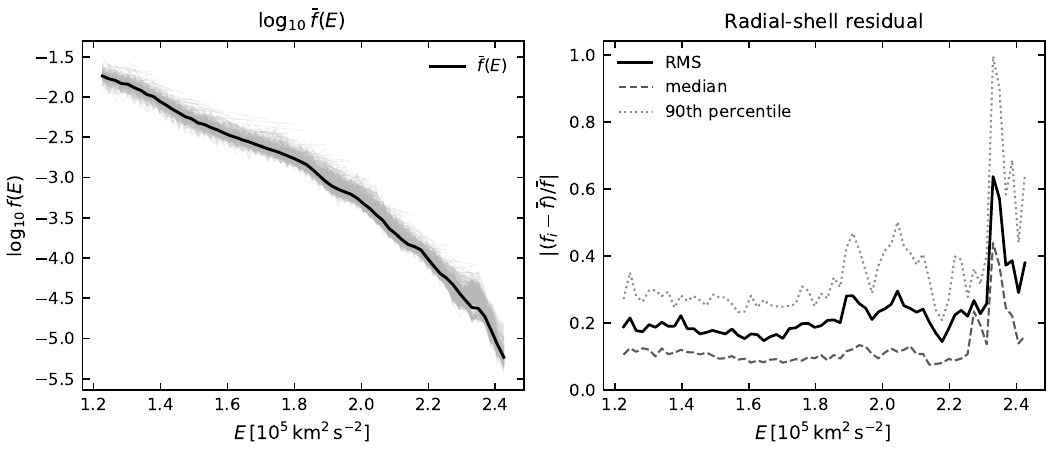}
    \caption{Inferred energy-only distribution function and radial-shell residuals for the Auriga halo 6 hydrodynamic run. Left: estimates of \(\log_{10} f_i(E)\) obtained independently in individual radial shells, shown as thin gray curves, together with the weighted shell-averaged distribution function \(\log_{10}\bar f(E)\) shown in black. Right: fractional residuals between the individual radial-shell estimates and the shell-averaged value, \((f_i-\bar f)/\bar f\), summarized as the RMS, median absolute residual, and 90th percentile absolute residual in each energy bin. This illustrates that the energy-only approximation captures the dominant radial-shell averaged behavior of the distribution function, while leaving residual shell-to-shell variations at the 
\(\sim 10\)--\(40\%\) level.}
    \label{fig:fE}
\end{figure}

We also find $\bar{f} (E)$ for each halo using the methods of Section~\ref{sec:General_Formalism},   to determine if an anisotropic phase space distribution describes the Auriga data significantly better than an isotropic 
distribution. Figure~\ref{fig:fE} shows the analogous energy-only reconstruction for the same representative halo shown in Figure~\ref{fig:fEL}. 
The gray curves show the estimates \(f_i(E)\) obtained independently in individual radial shells, while the black curve shows the weighted shell-averaged distribution \(\bar f(E)\). 
The right panel summarizes the corresponding shell-to-shell fractional residuals as a function of energy. In Table~\ref{tab:FEL_results}, we also present $N^E_{dof}$, $\chi^2_E$ and $(|\Delta f|/f)_{med.}^{E}$ for each halo, 
using $1000$ radial bins and $96$ energy bins.  We find typical values of $\chi^2_E /dof \sim 6-9$ (with 
$N^E_{dof} \sim 20000$), indicating that, statistically, the description of $f$ as a function of 
$E$ and $L$ is significantly better than as a function of $E$ alone.  
However, this statistical difference is largely driven by the fact that the number of particles 
in a typical energy and angular momentum bin used to compute $f(E,L)$ is far smaller than the number 
of particles in an energy bin used to compute $f(E)$.  As a result, the fractional uncertainties in $f(E,L)$ are 
typically far larger than the fractional uncertainties in $f(E)$ (that is, $|\delta f_i (E_j, L_k)|/f_i (E_j, L_k) \gg 
|\delta f_i (E_j)|/f_i (E_j)$), 
leading to a smaller $\chi^2/dof$ for $\bar{f}(E,L)$ than for $\bar{f}(E)$.  Although a phase space 
distribution which is a function of $E$ and $L$ is statistically a better fit to the numerical data 
than a distribution which is a function of $E$ alone, neither is a statistically good description of the 
data.  This is not surprising, since the underlying assumptions which would require the phase space distribution 
to be a function of $E$ and $L$ alone (spherical symmetry and equilibrium) are not exactly satisfied.

\begin{figure}
    \centering
    \includegraphics[width=1.0\linewidth]{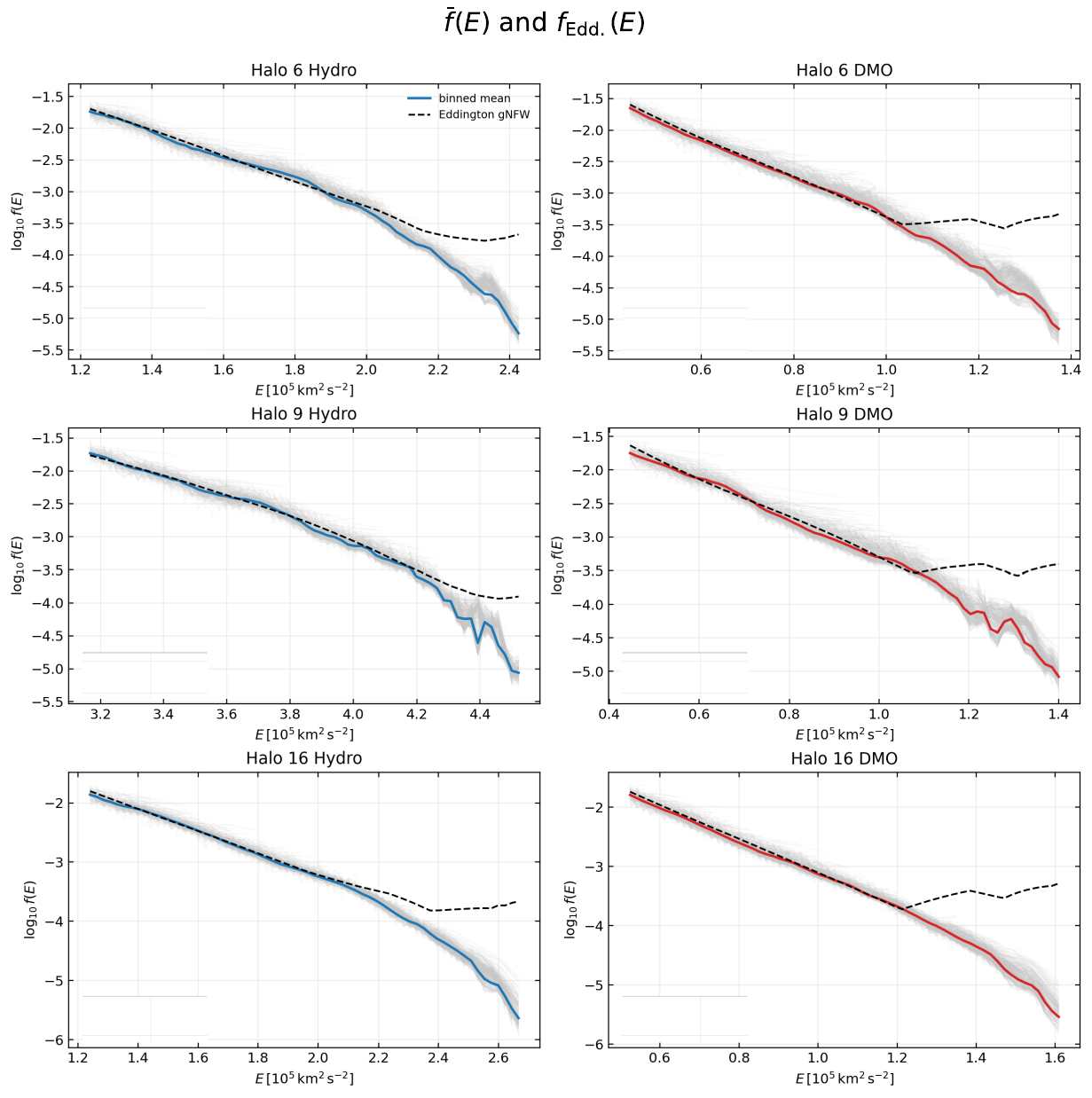}
    \caption{Inferred energy-only distribution function for all of the halos considered.  Estimates of \(\log_{10} f_i(E)\) obtained independently in individual radial shells are shown as thin gray curves, together with the weighted shell-averaged distribution function \(\log_{10}\bar f(E)\) shown as a solid line. $f_{Edd.} (E)$, obtained 
    from Eddington inversion, assuming a dark matter density profile given by the best fit gNFW profile, is shown 
    for each halo as a dashed line.}
    \label{fig:fE_all}
\end{figure}

Since neither $\bar{f}(E,L)$ nor $\bar{f}(E)$ are a statistically good description of the data, 
the relevant question is which of these two descriptions is a better approximation.  To answer this 
question, we again look at the fractional residuals.  
Comparing the values of $f_i (E_j)$ found in any radial bin to the $r$-independent distribution 
$\bar{f}(E_j)$ as described above, we find that the $\bar{f} (E)$ residuals are at the level of 
$(|\Delta f|/f)_{med.}^{E} \sim 0.15$.
Note that these fractional residuals are at a level comparable to $(|\Delta f|/f)_{med.}^{E,L} $.  
Moreover, rather than computing the median of the fractional residuals, we may instead compute the 
90th percentile fractional residual (denoted P90).  We then find 
$(|\Delta f|/f)_{P90}^{E} \sim (|\Delta f|/f)_{P90}^{E,L} \sim 0.4-0.6$.  That is, the $90\%$ of 
the fractional residuals in $(E,L)$ (or $(E)$, as relevant) bins are below $0.4 - 0.6$ in all of the 
halos we have considered.

From this, we can conclude that the description of the phase space distribution as a function of $E$ and 
$L$ is not a significant improvement over a description as a function of $E$ alone.  The fractional 
residuals in any particular radial bin are typically at the level of $\sim 15\%$, with $90\%$ of 
radial bins having residuals below $40-60\%$, depending on the halo.  Including an explicit dependence 
on $L$ in the phase space distribution $f(E,L)$ does not alter this picture significantly.

The choice of bins 
was made to minimize the associated binning error when analyzing Halo 6.  
To ensure that there is no 
bias, the same binning choice was also used on the other halos.

\begin{table}
    \centering
    \begin{tabular}{|c|c||c|c|c||c|c|c|}
    \hline
      Halo & baryons?  & $N^{E,L}_{dof}$  & $\chi_{E,L}^2/dof$ & $(|\Delta f|/f)^{E,L}_{med.}$ & $N^{E}_{dof}$ & $\chi_{E}^2/dof$ &  
       $(|\Delta f|/f)^{E}_{med.}$ \\
    \hline
      6 & Yes & $28149$ & $1.67$ & $0.15$ & $18682$ & $6.26$ & $0.12$ \\ 
      6 & No & $25872$ & $1.56$ & $0.148$ & $19783$ & $8.10$ & $0.15$ \\ 
      9 & Yes & $21948$ & $1.23$ & $0.131$ & $17317$ & $8.90$ & $0.13$ \\  
      9 & No & $20122$ & $1.24$ & $0.136$ & $19174$ & $9.51$ & $0.15$ \\ 
      16 & Yes & $39864$ & $1.87$ & $0.145$ & $19469$ & $7.51$ & $0.096$ \\ 
      16 & No & $39940$ & $1.90$ & $0.154$ & $19754$ & $8.29$ & 0.12 \\ 
    \hline
    \end{tabular}
    \caption{Table of halos considered, the number of degrees of freedom, $\chi^2/dof$ and 
    the median value of $(|\Delta f|/f$), for either the case in which $f$ is a function of $E$ and 
    $L$, or the case in which $f$ is only a function of $E$.  
     }
    \label{tab:FEL_results}
\end{table}

\subsection{A functional form for $f(E)$}
\label{sec:Function}

When the phase space distribution is isotropic, then $\rho(r)$ is related to $f(E)$ by eq.~\ref{eq:fE}, 
which is in the form of an Abel integral transform. The inverse Abel transform then yields 
the Eddington inversion formula~\cite{Eddington}
\bea
f_{Edd.}(E) &=& \frac{1}{\sqrt{8} \pi^2} \int_E^{\Phi (\infty)} \frac{d^2 \rho}{d\Phi^2} \frac{d\Phi}{\sqrt{\Phi - E}} .
\label{eqn:EddingtonInversion}
\eea
This expression is very useful because the density profile $\rho(r)$ can be directly estimated from 
observations of stellar kinematics, and in turn determines $\Phi (r)$.  With these observational inputs, 
one can then determine $f(E)$.  Indeed, in Ref.~\cite{Christy:2023mgs} it was noted that phase space 
distribution $f(E)$ obtained from publicly-available Via Lactea 2 simulated data (under the assumption of an 
isotropic phase space distribution) was similar to that which would have been obtained 
from the Eddington inversion formula, if one had only the observational inputs.

As a consistency check, in Figure~\ref{fig:fE_all} we plot $\bar f (E)$ obtained for each halo (as in 
Figure~\ref{fig:fE}), along with the $f_{Edd.}(E)$ obtained using eq.~\ref{eqn:EddingtonInversion}, assuming that the dark matter density 
profile $\rho(r)$ is of the gNFW form given by the best fit halo profile.  
We obtain the gravitational potential $\Phi (r)$ using eq.~\ref{eq:Phi}, 
where the enclosed mass is obtained from the 
simulated data (not the best fit gNFW profile), including baryons when relevant.
As we see, at low energy, $\bar f (E)$ is well-described 
by $f_{Edd.}(E)$.  This is not surprising, since the inverse Abel integral transform is unique; to the extent 
that $\rho(r)$ is well-described by a gNFW profile, and to the extent that the phase space density is 
well-described as a function of $E$ alone, it must be $f_{Edd.}(E)$.  The deviation of the $\bar f(E)$ from 
$f_{Edd.} (E)$ at large $E$ arises from the fact that the profile deviates from gNFW form and truncates near $R_{200}$, 
and from computational precision.

\section{Conclusions}
\label{sec:Conclusion}

We have considered the phase space distributions found in the suite of Auriga simulations, using 
both dark matter only runs and runs including baryons.  Under the assumption that the phase space 
distribution is spherically-symmetric and static, and describes particles moving in a central potential, 
principles of classical mechanics imply that the phase space distribution $f$ should be a function only of 
the energy  ($E$) and  angular momentum ($L$).  We have tested this conclusion in multiple Auriga simulated 
halos (in which the positions and velocities of the particles are known), and have found that $f$ can  
indeed be well-approximated as a function of $E$ and $L$ alone, without explicit dependence on $r$.  
Although this result is clearly not exact, the residuals are at the level of $\sim 15\%$, indicating the 
extent to which deviations from the assumptions of spherical symmetry and equilibrium affect the 
phase space distribution of the halo.
Since the dependence of the phase space distribution on $\vec{r}$ and $\vec{p}$ largely arises from the implicit 
dependence of $E$ and $L$ on these variables, our result implies that the dark matter velocity distribution 
is related to position in the halo in a particular manner encoded in the integrals of motion.

Moreover, we have found that the phase space distribution is equally well approximated as a function 
of $E$ alone, as might be expected from the principles of classical mechanics if the phase space 
distribution were isotropic.  In particular, if the phase space distribution is approximated as a function only 
of $E$, then the typical fractional residuals are similarly at the level of $\sim 15\%$; allowing an 
explicit dependence on $L$ does not cause a significant reduction of these residuals

Interestingly, these results hold (with similar residuals) in both dark matter only runs and runs 
including baryons.  This indicates that these conclusions should be robust when applied to real 
halos, such as that of the Milky Way, which are relevant for DM direct and indirect detection searches.

However, note that in this analysis we have obtained $f(E,L)$ from the distribution in $E$ and $L$ of 
the particles within radial shells.  Our result thus involves an element of spherical averaging, which is 
not present in the phase space distribution of an actual halo.  In particular, GAIA stellar kinematic data has been 
used as a tracer of the bulk motion of dark matter streams, showing important deviations from a 
spherically-symmetric, static and isotropic phase space distribution~\cite{Necib:2018iwb}. 
We estimate the impact of these effects on the 
phase space distribution to be at the level of $\sim 15\%$ once spherically averaged, but the localized impact can 
be even larger.  This can be especially important for direct detection searches, where one is interested 
in the velocity distribution at the location of the Earth~\cite{Vogelsberger:2008qb,Freese:2012xd}.

Although we have seen that one does not lose significant accuracy in modeling the spherically-averaged 
phase space distribution as being isotropic (that is, independent of $L$), the impact may be more substantial 
when applied to directional dark matter direct detection (for a recent review, see Ref.~\cite{Vahsen:2020pzb}).  
In that case, anisotropies in the dark matter 
velocity distribution within the solar system are more important, which may require including the 
dependence of the phase space distribution on $L$.

We have used the Auriga suite of simulations, which provide simulated data for  Milky Way-sized 
dark matter halos.  There are subhalos within these MW-sized halos, but it is difficult to perform 
a similar analysis on the subhalos because there are so few particles that it is difficult to accurately 
estimate the density profile and  gravitational potential, which are crucial inputs.  It would be interesting 
to perform a similar analysis on simulations with high-resolution at the scale of subhalos that are the 
size of dwarf spheroidal galaxies, to see if these results still hold.

{\bf Acknowledgements}  We are grateful to Taylor Herbert for collaboration at the initial stage of this project, and to Nassim Bozorgnia and Isabelle Goldstein for useful discussions. JK and LES wish to acknowledge the Center for Theoretical Underground Physics and Related Areas (CETUP*), the Institute for Underground Science at Sanford Underground Research Facility (SURF), and the South Dakota Science and Technology Authority for hospitality and financial support, as well as for providing a stimulating environment. JK is supported in part by DOE grant DE-SC0010504B. LES is supported in part by DOE grant DE-SC0010813. The authors acknowledge the use of OpenAI Codex to assist with code and figure development for this paper. All generated results were verified and the authors take responsibility for all results presented in this paper.

\end{document}